\documentstyle[floats,multicol,aps,epsf,prb]{revtex}

\begin{document} 
\draft
 \twocolumn[\hsize\textwidth\columnwidth\hsize\csname @twocolumnfalse\endcsname

\title{ On the Abrikosov transition in disordered superconducting films.}
\author{A.V. Lopatin}
\address{Department of Physics, Rutgers University, Piscataway, 
New Jersey  08855}
\date{\today}
\maketitle

\begin{abstract}

 We consider a disordered  superconducting film in the magnetic field close
to the upper critical field. Assuming that the phase transition to the glassy  
superconducting state exists and it is of the second order we show that it is
exactly described by a zero-dimensional replica model having
a form of one describing the phase transition in the SK 
(Sherrington-Kirkpatrick) spin-glass model. The dependence of the magnetic 
moment on the external magnetic field is shown to be more smooth than the 
Abrikosov mean field result for the clean case. While the total 
magnetic moment of the film close to the phase transition is diamagnetic, 
the part of it emerging due to the phase transition is paramagnetic, 
this shows tendency to the paramagnetic Meissner effect.      
 
\end{abstract}

  \vskip2pc]

 A magnetic field  penetrates type-II superconductors forming the Abrikosov 
lattice of magnetic vertices \cite{abr}. Obviously, formation of a purely periodical 
structure is possible only in homogeneous materials. The influence of disorder
on the structure of the Abrikosov lattice was first  considered in Ref.\cite{larkin} 
assuming that the random force acting on the lattice is uncorrelated with the  
lattice displacements and it was shown that  the long range order is destroyed by 
inhomogeneity. Later this problem was considered in Ref.\cite{giam} using more 
complicated ( but still not rigorous) methods  where the result of absence of the 
long range order was confirmed and the correlation functions were analyzed more 
carefully. It is expected theoretically and it was seen experimentally \cite{exp}  
that this system should have glassy properties. Therefore a  systematic treatment 
of this problem  is very complicated and is absent at the  present  time. 
     
 We consider  a disordered  superconducting film in the external  magnetic field 
close to the upper critical field $H_{c2}$ and show that the phase transition to the 
glassy superconducting state is exactly described by a zero dimensional 
replica model having a  form of one describing the phase transition in the 
infinite-range SK  spin-glass model. To obtain this result we
assume that the phase transition exists and it is of the second order. 
The dependence of the magnetic moment $M$ on the external magnetic field 
(or on temperature) following from this theory is smooth, only the first 
derivative $dM/dH$ ($dM/d T$) has a kink in contrast with the Abrikosov mean 
field result for the clean case. The part of the magnetic moment emerging due 
to the transition to the glassy superconducting state is paramagnetic, i.e. 
the system exhibits  tendency to the paramagnetic Meissner effect.

Our starting point is the Ginzburg-Landau free energy density
\begin{equation}
{\cal F}=\phi^*\Big(-{{D^2}\over{2m^*}}+u(r)+a\Big)\phi+
{b\over 4}\,\,\phi^*\phi^*\phi\phi
+{{{\bf B}^2}\over{8\pi}}-{{{\bf B} {\bf H}_0}\over{4\pi}} , \label{gl}
\end{equation}
where $D={\bf \nabla}+i e^* {\bf A}/c,\,\,e^*=2|e|,$ and ${\bf H}_0$ is an  external 
magnetic field which is perpendicular to the film. The order parameter 
field $\phi$ in (\ref{gl}) is supposed to be normalized such that $a=T-T_c,$
where $T_c$ is the mean field critical temperature. 
The width of the film $d$ is supposed to be smaller 
than the  coherence length. The potential $u(r)$ models the disorder in the 
system and it is assumed to be Gaussian with the correlator  
$\langle u({\bf r}_1) u({\bf r}_2)\rangle=\sqrt{u}\,\,\delta(r_1-r_2).$
This term emerges because in the random system the coefficient $a=T-T_c$
becomes a random quantity. In fact all the coefficients in (\ref{gl}) contain
randomness, but apparently, close to the phase transition the most relevant
effect comes from randomness of the coefficient $a$ which we consider.

Using the replica trick we can average (\ref{gl}) over $u(r)$ getting
the effective functional

\begin{eqnarray}
{\cal F}\,n=\sum_\alpha \phi_\alpha^*\Big(-{{D^2}\over{2m^*}}+a\Big)\phi_\alpha-
{u\over 2}\sum_{\alpha,\beta}\phi_\alpha^*\phi_\beta^*\phi_\alpha\phi_\beta 
\nonumber  \\
+{b\over 4}\sum_\alpha \phi_\alpha^*\phi_\alpha^*\phi_\alpha\phi_\alpha
+{{{\bf B}^2}\over{8\pi}}-{{{\bf B} {\bf H}_0}\over{4\pi}}, \label{l}
\end{eqnarray}
where $\alpha,\beta$ are the replica indexes taking values from $1$ to $n$.
In the lowest Landau level approximation the fluctuations of the magnetic 
field can be integrated out giving the renormalization of the interaction 
term \cite{affleck}
\begin{eqnarray}
{\cal F}^\prime n=\sum_\alpha \phi_\alpha^*(-{{D^2}\over{2m^*}}+a)\phi_\alpha-
{u\over 2}\sum_{\alpha,\beta}\phi_\alpha^*\phi_\beta^*\phi_\alpha\phi_\beta 
\nonumber \\
+{v\over 4}\sum_\alpha \phi_\alpha^*\phi_\alpha^*\phi_\alpha\phi_\alpha, \label{fun}
\end{eqnarray}
where  $v=b-(e^*/2mc)^2/2\pi$ and ${\cal F}^\prime={\cal F}+{\bf H}_0^2/8\pi.$ 
We shall take the gage
$A_x=-Hy$ and use the  Landau functions to diagonalize the quadratic part of  the 
functional (\ref{fun})
\begin{equation}
\phi_p={1\over{\pi^{1/4} \sqrt{L_x l}}}\,\,e^{-{1\over 2}(y/l-pl)^2+ipx},
\end{equation}
where $l=\sqrt{c/e^*H},$ $L_x$ is the size of the film in the $x-$ direction, 
and $p$ is the 
momentum. In this basis the functional (\ref{fun}) becomes
\begin{eqnarray}
{\cal F}^\prime n=
\sum_{ p_1+p_2 =p_3+p_4}K_{p_1 p_2 p_3 p_4} 
\Big[-{u\over 2}
\sum_{\alpha,\beta}\phi_{\alpha  p_1}^*\phi_{\beta  p_2}^*
\phi_{\alpha p_3}\phi_{\beta p_4} \nonumber \\
+{v\over 4}\sum_\alpha \phi_{\alpha p_1}^*
\phi_{\alpha p_2}^*\phi_{\alpha p_3}\phi_{\alpha p_4}\Big]
+T_c \sum_{p,\alpha}\delta\,\, \phi_{\alpha  p}^*\phi_{\alpha p}, \label{lll}
\end{eqnarray}
where $\delta=(e^*H/mc+a)/T_c$ and the kernel $K$ is
\begin{equation}
K_{p_1 p_2 p_3 p_4}={1\over{L_x l\sqrt{2\pi}}} e^{-l^2[(p_1-p_2)^2+(p_3-p_4)^2]/4}.
\end{equation}
Note that the Lagrangian (\ref{lll}) is translationally invariant in the momentum $p.$
In the clean case this invariance is broken when the Abrikosov lattice is formed.
 But in the disordered case which we consider we expect that all averaged quantities 
are translationally invariant in $p.$ Therefore the averaged Green function 
\begin{equation}
G_{\alpha,\beta}(p)=\langle\phi_{\alpha p}^*\phi_{\beta p}\rangle
\end{equation}
should not depend on $p.$ The functional (\ref{lll}) as usually can be presented as an 
expansion in the total Green function $G_{\alpha,\beta}(p)$. Deriving this effective 
functional one can
perform all  momentum integrations in each diagram because  the momentum dependence 
of the Green function is trivial, and the resulting free energy functional contains 
$G_{\alpha,\beta}$ depending only on the replica indexes. Keeping the terms up to 
the forth order in $G$ we get
\begin{eqnarray}
&F^\prime n&=N\, T_c\biggl[-Tr\log{G}+\delta\,\, Tr G+\kappa
\sum_{\alpha} (G_{\alpha,\alpha})^2
-\theta\,\, Tr G^2  \nonumber \\
&-&{{\kappa^2}\over 4}\sum_{\alpha,\beta} (G_{\alpha,\beta})^4-{{\theta^2}\over 2}
Tr G^4+
\kappa\theta\sum_\alpha \Bigl([G^2]_{\alpha,\alpha}\Bigr)^2 \biggr], \label{repl}
\end{eqnarray}
where the  dimensionless coupling constants  are  $\kappa=v/4\pi T_c l^2 d,$ 
$\theta=u/ 4\pi T_c l^2 d, $ and $N=e^* H_0 L_x L_y/2\pi c$ is the number of quantum 
states at the lowest Landau level. Let us first analyze the model (\ref{repl}) and 
then we will discuss how the results are affected by the high order terms neglected 
in (\ref{repl}). Variation of (\ref{repl}) with respect to $G_{\alpha,\beta}$ gives 
the equation defining $G_{\alpha,\beta}.$ In the normal phase the solution should be 
replica symmetric $G_{\alpha,\beta}=g\, \delta_{\alpha,\beta}$ because
the model (\ref{fun}) is symmetric under the global $U(1)$ transformations  
within each replica. Variation of (\ref{repl}) with respect to 
$g$ gives
\begin{equation}
-{1\over g}+\delta+2(\kappa-\theta)g+(4\kappa\theta-\kappa^2-2\theta^2)g^3=0 
\label{delta}
\end{equation}
 In the glassy superconducting state we expect a non-replica-
symmetric solution for $G_{\alpha,\beta}$ to appear.  The glass transition corresponds 
to the instability of the replica-symmetric solution. Presenting 
$G_{\alpha,\beta}$ as a sum of diagonal and off-diagonal 
parts $G_{\alpha,\beta}=g\,\delta_{\alpha,\beta}+Q_{\alpha,\beta}$
and expanding (\ref{repl}) to the second orderer in $Q$ we get
\begin{equation}
N T_c\sum_{\alpha,\beta}Q_{\alpha,\beta}\Big[{1\over{2 g^2}}-\theta +g^2 
\theta\,(2\kappa-3\theta)\Big] 
Q_{\alpha,\beta}.
\end{equation}
 At the glass transition this term  becomes zero. This gives the value of $g$ 
at the phase transition
\begin{equation}
g_c^2={1\over {2\theta}}{{-1+\sqrt{7-4\kappa/\theta}}\over {3-2\kappa/\theta}}.
\label{gc}
\end{equation}
 We see that within the approximation which we used the second order phase transition 
is possible only when the disorder is strong enough $\theta/\kappa>4/7.$ 
The value of $\delta$ at the phase transition is determined by  (\ref{delta}) 
\begin{equation}
\delta_c={1\over {g_c}}-2(\kappa-\theta)g_c-(4\kappa\theta-\kappa^2-2\theta^2)g_c^3=0. 
\end{equation}
We will be interested in the region close to the glass transition where  we can 
simplify (\ref{repl}) expanding in $Q.$ Taking $g=g_c+g^\prime,\,\delta=\delta_c+
\delta^\prime,$
expanding in $Q,g^\prime$ and eliminating $g^\prime$ we get the following effective 
free energy functional
\begin{eqnarray}
n F_Q =N T_c\biggl[\delta^\prime Tr\, Q^2-{\gamma\over 3} Tr\, Q^3 -{\alpha_1\over 4}
\sum_{\alpha,\beta}(Q_{\alpha,\beta})^4 \nonumber \\
-{\alpha_2\over 4}\sum_{\alpha} ([Q^2]_{\alpha,\alpha})^2
+{\alpha_3\over 4} Tr Q^4\biggr], \label{model}
\end{eqnarray} 
where
\begin{eqnarray}
\delta^\prime&=&{\theta\over{g_c\kappa}}\,{{\sqrt{7-4\kappa/\theta}} 
\over{1+g_c^2\,\theta\,(4-3\kappa/2\theta)}}\,(\delta-\delta_c) \label{coef1} \\
\gamma&=&g_c\Big(6\,\theta^2+{1\over {g_c^4}}\Big) \label{coef2}     \\
\alpha_1&=&\kappa^2. \label{coef3}
\end{eqnarray}
We do not write the expressions for the  coefficients $\alpha_2,\alpha_3$ 
because they do not affect the results in the leading order \cite{Fischer93}. 
The model (\ref{model}) is standard in the spin-glass theory, 
it describes the phase transition in the SK spin-glass model
(see for example Ref.\cite{Fischer93}). 
The solution is known to have the form of the Parisi matrix which is 
parameterized by the function  $q(x),\,0<x<1,$ after taking the limit $n \to 0.$
For the model (\ref{model}) the function $q(x)$ is
 \begin{eqnarray}
q(x)&=& x\,\,{{q_m}\over{x_m}}, \;\;\;\; x<x_m, \\
q(x)&=& q_m, \;\;\;\;\;  x_m<x<1,  
\end{eqnarray}
where $x_m=-{{3\alpha_1}\over{\gamma^2}} \delta^\prime$ and
$q_m=-{{\delta^\prime}\over \gamma}.$

 Now let us  analyze the validity of the obtained results: According
to (\ref{gc}) the Green function $G_{\alpha,\beta}$ is of the order of 
$1/\sqrt{\theta}.$
Therefore the high order corrections  to (\ref{repl}) are of the same order 
or less than the terms that were kept. It means that the values of the 
coefficients (\ref{coef1},
\ref{coef2},\ref{coef3}) are correct only by the order  of magnitude.
Also Eq.(\ref{gc}) can be used only as an estimation, therefore
the existence of the second order phase transition in fact is an 
assumption. 
Nevertheless, the form of the functional (\ref{model}) is correct because
it contains all terms up to the forth order allowed by the replica symmetry of the 
original model and the expansion in $Q$ is always possible close to the phase 
transition. Therefore the phase transition is effectively described by a 
zero-dimensional model 
(not including  the  replica indexes).
While this is typical for the infinite range spin-glass models it is a very surprising 
result for a real physical system.

  The magnetic moment of the system can be written  as
\begin{equation}
M=-{{d F^\prime}\over{d H_0}}= M_{fl}+M_Q, \label{magnmom}
\end{equation}
where $M_Q$ is the  part of the magnetic moment emerging due to formation of 
the glass state
\begin{equation}
M_Q=-{{ d F_Q}\over{d H_0}},
\end{equation}
and $M_{fl}$ is the part which can not be found from the effective functional 
(\ref{model}) but it should not  have any peculiarities  at the phase transition. 
In the normal state $M_{Q}=0$ and in the glass state from (\ref{model}) we have
\begin{equation}
M_Q\sim -{{{TrQ^2}\over n}}_{n\to 0} \sim  \delta^{\prime 2}. \label{mq}
\end{equation}
 We see that the dependence of the magnetic moment on the external magnetic field 
is smooth only the susceptibility dependence $d M/dH_0$ has a kink at the phase 
transition. Since the temperature enters the functional (\ref{lll}) in the same 
way as  the magnetic field, analogously we conclude that  the entropy is a smooth 
function of the temperature and the specific heat dependence on 
the temperature  has a kink. 
For a comparison let us note  that in the pure case the Abrikosov mean field theory 
predicts more singular behavior: the magnetic moment and specific heat dependences 
should have  kinks at the phase transition.   
   
  Within the lowest Landau level approximation the total magnetic moment (\ref{magnmom})
is always negative, this is clear from (\ref{lll}) because $\langle\phi^*_{\alpha,p}
\phi_{\alpha,p}\rangle >0.$ But the part of the magnetic moment emerging due to the
transition to the glassy superconducting state is positive according to (\ref{mq}).
It means that the system has tendency to exhibit the paramagnetic Meissner effect.    

 This theory can be applied for experiments on Nb superconducting films
in magnetic fields close to $H_{c2}.$ To the best of our knowledge the
region close to the phase transition was not yet investigated.
Our theory predicts classical critical exponents for dependences
of magnetization and specific heat on  temperature or magnetic field. 
The dependence of the magnetization $M$ on the temperature
or magnetic field is smooth, only the first derivatives $dM/d T,\,\,\,
d M/d H_0$ should have  kinks.
 Since our theory predicts that the  magnetic moment emerging due to
the phase transition is paramagnetic, experimentally one would expect that  
decreasing the temperature, the magnetization first decreases
due to the effect of the diamagnetic fluctuations, but then increases
bellow the phase transition.

 It is interesting to mention the experiment on the superconducting disks \cite{exp}
where the phase transition region was investigated in details.
The results of this experiment qualitatively agree with our theory    
,but the thickness of the disks that
were used is larger then the coherence length, therefore our theory
can hardly be directly applied.  
  
I am grateful to Elihu Abrahams, Thierry Giamarchi, Lev Ioffe, Gabriel Kotliar and A.I. Larkin
for very useful discussions.

\end{document}